\documentclass[11pt]{article}

 \usepackage{latexsym,amsfonts,amsmath,amssymb,wasysym,enumerate,epsfig}
\usepackage{theorem}
\usepackage{mathrsfs}
\usepackage{tikz}
\usepackage{subfig}

\usepackage{color}

\numberwithin{equation}{section}

\newcommand{\mb}[1]{\quad\mbox{#1}\quad}
\newcommand{\beq}{\begin{equation}}
\newcommand{\eeq}{\end{equation}}
\newcommand{\beu}{\begin{equation*}}
\newcommand{\eeu}{\end{equation*}}
\newcommand{\bea}{\begin{eqnarray}}
\newcommand{\eea}{\end{eqnarray}}
\newcommand{\beano}{\begin{eqnarray*}}
\newcommand{\eeano}{\end{eqnarray*}}
\newcommand{\bmx}{\begin{pmatrix}}
\newcommand{\emx}{\end{pmatrix}}

\newcommand{\eps}{\varepsilon}

\newcommand{\PP}{\mathbb{P}}

\newcommand{\LL}{{\mathbb L}}

        \def\cO{{\cal O}}

    \def\cZ{{\cal Z}}

\newcommand{\wh}[1]{{\widehat{#1}}}
\newcommand{\wt}[1]{{\widetilde{#1}}}
\newcommand{\wb}[1]{{\overline{#1}}}
\newcommand{\nonu}{\nonumber\\}

\newcommand{\llangle}{\langle}
\newcommand{\rrangle}{\rangle}

\newcommand{\bP}{{\boldsymbol{P}}}

\advance\textheight by \topskip
\begin{document}
\setcounter{page}{0}
\pagestyle{empty}
%
%
\begin{center}

 {\LARGE  {\sffamily Matrix product solution to a 2-species TASEP \\[1ex]
 with open integrable boundaries} }\\[1cm]

\vspace{10mm}
  
{\Large 
 N. Crampe$^{a}$\footnote{nicolas.crampe@umontpellier.fr},
 M. R. Evans$^{b}$\footnote{mevans@staffmail.ed.ac.uk},
 K. Mallick$^{c}$\footnote{kirone.mallick@cea.fr},
 E. Ragoucy$^{d}$\footnote{eric.ragoucy@lapth.cnrs.fr}
 and M. Vanicat$^{d}$\footnote{matthieu.vanicat@lapth.cnrs.fr}}\\[.41cm] 
{\large $^a$ Laboratoire Charles Coulomb (L2C), UMR 5221 CNRS-Universit{\'e} de Montpellier,
Montpellier, F-France.}
\\[.42cm]
{\large $^b$ SUPA, School of Physics and Astronomy, University of Edinburgh,
Peter Guthrie Tait Road, Edinburgh EH9 3FD, UK}
\\[.42cm]
{\large $^c$ Institut de Physique Th{\'e}orique, CEA Saclay, F-91191 Gif-sur-Yvette, France}
\\[.42cm]
{\large $^{d}$ Laboratoire de Physique Th{\'e}orique LAPTh,
 CNRS and Universit{\'e} de Savoie.\\
   9 chemin de Bellevue, BP 110, F-74941  Annecy-le-Vieux Cedex, 
France. }
\end{center}
\vfill

\begin{abstract}
We present an explicit representation for the matrix product ansatz for some two-species TASEP with open boundary conditions. 
The construction relies on the integrability of the models, a property that constrains the possible rates at the boundaries. 
The realisation is built on a tensor product of copies of the DEHP algebras. 
Using this explicit construction, we are able to calculate the partition function of the models. 
The densities and currents in the stationary state are also computed. 
It leads to the phase diagram of the models. Depending on the values of the boundary rates, we obtain for each species shock waves, 
maximal current, or low/high densities phases.
\end{abstract}

\vfill\vfill

\rightline{LAPTh-032/16\qquad}
\newpage
\pagestyle{plain}
\section{Introduction}
The totally asymmetric exclusion process (TASEP) has proved itself to
be a paradigmatic model for non-equilibrium, current-carrying states
\cite{CKZ,BE}.  The model comprises particles hopping stochastically
along a one-dimensional lattice with hard-core exclusion interactions.
The version with open boundary conditions, where particles enter and
leave at the boundaries, has revealed the existence of boundary
induced phase transitions \cite{Krug,DEHP,SD}. An exact solution of
the stationary state is provided by a matrix product Ansatz
\cite{DEHP,BE} which allows the stationary density profiles and all
equal-time correlation functions to be computed. The combinatorial
nature of this solution has been explored in various works
\cite{Angel,Brak,DS,FM1,FM2}.  Moreover, recent works have explored
the integrability of the model and allowed the spectrum of the Markov
matrix governing the dynamics to be established via the Algebraic
Bethe Ansatz \cite{dGE,Crampe}.

The generalisation to several species of particles has been under
intense study, initially motivated by the case of second-class
particles.  Second-class particles hop forward when there is a vacancy
immediately in front of them but may be overtaken by a first-class
particle immediately behind them, in which case they hop back one
lattice site.  Second-class particles have proven to be a useful tool
in tracking the positions of microscopic shocks in the system
\cite{Ferrari,FKS}.  In the case of periodic boundary conditions the
exact stationary state of a system of first and second class particles
has been established in a matrix product formulation \cite{DJLS}.
Inspired by work in the probabilistic literature which invoked a
queueing interpretation \cite{FM1,FM2}, the solution was later
generalised to  a multi-species hierarchy \cite{EFM,MMR,PEM,AAMP1}. The
resulting solution  is constructed from tensor products of   the
fundamental matrices used in the two-species solution.  The Bethe
integrability of the multi-species problem  has been established
through co-ordinate Bethe Ansatz for the two-species case
\cite{AB,DE99,Cantini}  and  by Algebraic Bethe Ansatz for the
multi-species case \cite{AKSS,AAMP2}.

Various examples of open boundary conditions with two or more species
of particles  have been considered initially motivated by boundary-induced symmetry breaking transitions
\cite{EFGM1}.  Matrix product solutions have been
found in the two-species case where the stationary current of
second-class particles vanishes, which occurs under certain symmetric
boundary conditions \cite{EFGM2,DE} or when the second-class particles
are constrained to remain in the system through semi-permeable
boundaries \cite{Arita0,Arita1,Uchiyama,ALS1,ALS2}.

However, the case of general open boundary conditions for  a system of
first and second-class particles has remained an elusive
problem. Recently  progress was made in identifying classes of
boundary conditions for which the system is integrable, in the sense
of the Algebraic Bethe Ansatz \cite{CMRV}. In that work the algebraic
structure of the stationary state was worked out in a particular case
of integrable boundary conditions. 

In the present work we consider the manifold of integrable boundary
conditions established in \cite{CMRV} and show that these boundary
conditions lead to a non-trivial phase diagram with phases that
manifest non-zero currents of the second-class particles. We find an
explicit solution for the stationary state in terms of tensor products
of the fundamental matrices which appear in the two-species solution
for periodic boundary conditions.  Interestingly, the stationary state
has various factorisation properties and these allow us to compute
exactly the partition function for the system.  The proof of the
stationary state is simplified by the factorisation properties and
allows us to make the connection with techniques used in the study of
integrable systems such as Zamolodchikov-Faddeev and
Ghoshal-Zamolodchikov equations.

The paper is organised as follows.  In section 2 we review the
integrable boundary conditions established in \cite{CMRV} and derive
the corresponding phase diagrams.  We then present in section 3 the
tensor product solution for the stationary state and present various
factorisation properties, and  use this solution to compute the
partition function for various cases. In section 4  we prove  that the
tensor product state is stationary under the Markov matrix that
generates the dynamics.  In section 5,  we elucidate the connection to
techniques of Algebraic Bethe Ansatz. We conclude in section 6  with
an overview of open problems. Finally, supplementary technical details are 
presented in appendices.

\section{Definition of the models and Phase Diagrams}

The two-species totally asymmetric exclusion process (2-TASEP)
is a stochastic dynamical system,  defined on  a
one-dimensional lattice with $L$ sites  in contact with two  boundary
reservoirs,  where each site
$i=1,\ldots,L$ can be in one of three states $\sigma_i=0$, $1$ or $2$.
State $0$ may be considered as an empty site or hole. State $1$
corresponds to a first class particle, and state $2$ corresponds to a
second class particle.  In the bulk, at each pair of nearest neighbor
sites, the rates of exchange are
\begin{equation}
 1\ 0 \xrightarrow{1} 0\ 1 \quad,\qquad 2\ 0  \xrightarrow{1}
 0\ 2\quad,\qquad 1\ 2  \xrightarrow{1} 2\ 1\;.
\end{equation}
(We remark that various other conventions for the labelling of the three particle states have been employed in the literature e.g. \cite{EFGM1,EFGM2}.)
The sites $1$ and $L$ are in contact with boundary reservoirs and particles are
exchanged at  different rates at the boundaries. For generic values of
these  boundary rates, the system is not integrable (in contrast with
the 1-species TASEP, which is integrable for arbitrary  boundary
rates);  finding an exact solution looks hopeless. However,  in
\cite{CMRV},   using a systematic procedure, all  possible  boundary
rates for  the 2-species TASEP that preserve   integrability were
classified. Amongst  such  models, some  had  been studied  earlier:
the first open two-species matrix product solutions were derived in \cite{EFGM2}; in \cite{DE} the boundary conditions for which
the stationary state may be expressed using the matrices
$D$, $E$, $A$ of \cite{DEHP,DJLS} were deduced;
in \cite{ALS1,ALS2} the restricted class of
{\it semi-permeable} boundaries, in which  second class
particles can neither enter nor leave the system was studied.  
In all of these cases  a matrix product
representation of the stationary state was found 
involving the quadratic algebra  used   by Derrida, Evans, Hakim and
Pasquier \cite{DEHP} in   their exact  solution of the 1-species
exclusion process with open boundaries. 

  In the present work, we construct a matrix Ansatz for  integrable
  2-TASEP with open boundaries that allow all  species of
  particles to  enter and  leave the system. The algebraic structures
  required  will be much more involved  than the fundamental quadratic
  algebra of  \cite{DEHP}.  

 We shall study  two classes of  2-species TASEP models with the following
 boundary  rates
\begin{eqnarray}
\begin{array}{c c c}
&\text{left boundary}\hspace{1cm}&\text{right boundary}\\
& 2 \xrightarrow{\hspace{2mm}\alpha\hspace{2mm}} 1\hspace{1cm} & 2 \xrightarrow{\hspace{2mm}\beta\hspace{2mm}} 0\\
(P_1):& 0 \xrightarrow{\hspace{2mm}\alpha\hspace{2mm}} 1\hspace{1cm}&  1 \xrightarrow{\hspace{2mm}\beta\hspace{2mm}} 0\\
& 0 \xrightarrow{1-\alpha} 2\hspace{1cm}&   1 \xrightarrow{1-\beta} 2
 \end{array}\label{P1}
\end{eqnarray}
or 
\begin{eqnarray}
\begin{array}{c  c c}
& \text{left boundary}\hspace{1cm}&\text{right boundary}\\
 &2\xrightarrow{\hspace{2mm}\alpha\hspace{2mm}} 1\hspace{1cm} & 2 \xrightarrow{\hspace{2mm}\beta\hspace{2mm}} 0\\
(P_2): &0 \xrightarrow{\hspace{2mm}\alpha\hspace{2mm}} 1\hspace{1cm}&  1 \xrightarrow{\hspace{2mm}\beta\hspace{2mm}} 0\\
 &0 \xrightarrow{1-\alpha} 2\hspace{1cm}& 
 \end{array}\label{P2}
\end{eqnarray}
 Hereafter, the  two  different models will be denoted by $(P_1)$ and
 $(P_2)$.  
Note that in  the classification of \cite{CMRV} the left
 boundary conditions were referred to as $L_2$ and the right hand
 boundary conditions for  $(P_1)$ or $(P_2)$ were referred to as $R_2$ and
 $R_3$ respectively. It is a simple matter to translate our results for $(P_2)$ to the case
of right boundary $R_2$ and left boundary $L_3$. The final case of
right boundary $R_3$ and left boundary $L_3$ leaves the stationary state devoid of holes and thus reduces to a one-species TASEP.

The physical interpretation of the boundary conditions is as follows.
In both models $(P_1)$, $(P_2)$ the left-hand boundary conditions correspond to a boundary reservoir containing only first and second class particle
with densities $\alpha$ and $1-\alpha$ respectively,  with no holes.
In model ($P_1$) the right-hand boundary conditions correspond to a reservoir containing second-class particles and holes
with densities $1-\beta$ and $\beta$ respectively, with no first-class particles.
In model ($P_2$) the right-hand boundary conditions correspond  to a reservoir containing first-class particles and holes
with densities $1-\beta$ and $\beta$ respectively, with no second-class particles.

 The 2-TASEP is  a finite   Markov process that reaches a unique
 steady-state in the long time limit,  in which   each configuration
 has the  stationary probability  (or weight)  $P(
 \sigma_1,\sigma_2,\dots,\sigma_L)$.  The  column-vector $\bP$ of
 length $3^L$, whose components are the probabilities $P(
 \sigma_1,\sigma_2,\dots,\sigma_L)$, satisfies the stationary master equation
\begin{equation}\label{master_equation}
 M^{(3)}\bP=0,
\end{equation}
where $M^{(3)}$ is the $3^L \times 3^L$ Markov matrix for the 2-TASEP
system.  Finding the  steady state thus amounts to solving a linear
system that grows  exponentially with the size of the system.
Basically, the matrix product representation of the  stationary
weights, based on integrability, will  allow us to reduce this
exponential complexity to a polynomial computation.
 

\subsection{Phase diagrams}

 The stationary state of the exclusion process can exhibit different
 qualitative features and different analytical expressions 
 for macroscopic quantities in the infinite   size limit,  $L \to \infty$.
 The system  is said to exhibit various phases,  that depend on the
 values of the boundary exchange rates.  These different phases  can
 be discriminated by  the values of the currents and  by the shapes of
 the density profiles. More refined features, such as correlations
 length  or even dynamical behaviour,  can  even lead us to define
 subphases (see \cite{BE}  for details and references).  The  phase
 diagram of the one-species TASEP has been  well-known for a long
 time; first determined using  a mean-field  approximation
 \cite{Krug,DDM}, it was rigorously established  and precisely
 investigated after  the finding of the exact solution
 \cite{DEHP,SD,BE}.  We recall that  the dynamical rules of the
 one-species TASEP  are given by \beq
\begin{array}{c c c}
\text{left boundary}\hspace{1cm}&\hspace{1cm}\text{bulk}\hspace{1cm}&\hspace{1cm}\text{right boundary}\\[1ex]
0 \xrightarrow{\hspace{2mm}\alpha\hspace{2mm}} 1\hspace{1cm} &1\ 0 \xrightarrow{\hspace{2mm}1\hspace{2mm}} 0\ 1& 1 \xrightarrow{\hspace{2mm}\beta\hspace{2mm}} 0
\end{array}
\eeq
The phase diagram is determined by the behavior of the stationary current $J$ 
and bulk density of particles in the limit $L\to\infty$ \cite{BE}. The different phases
 are detailed in table \eqref{phase_diag_1-TASEP}.

\begin{eqnarray} \label{phase_diag_1-TASEP}
\begin{array}{c|c|c|c} 
\mbox{Region}& \mbox{Phase} & \mbox{Current } J & \mbox{Bulk density}  \\ \hline
\rule{0pt}{3.54ex}\alpha<\beta,\,\alpha<\frac{1}{2} & \mbox{Low-density (LD)} & \alpha(1-\alpha) & \alpha \\[1ex]
\beta<\alpha,\,\beta<\frac{1}{2} & \mbox{High-density (HD)} & \beta(1-\beta) & 1-\beta \\[1ex]
\alpha>\frac{1}{2},\,\beta>\frac{1}{2} & \mbox{Maximal current (MC)} & \frac{1}{4} & \frac{1}{2}
 \end{array}
\vspace*{1ex}
\end{eqnarray}

The  phase diagrams of the  $(P_1)$ and $(P_2)$ models can be
determined rigorously  without having to compute exactly the steady-state
probabilities.  Indeed,  the various phases  of these two
species models can extracted from the knowledge of the one-species
TASEP phase diagram, by  using an identification procedure formalised
in \cite{ALS2}.  Then,   2-TASEP
models $(P_1)$ and $(P_2)$ can both  be mapped to  the
one-species TASEP model using  two possible identifications:
\begin{enumerate}
\item One can identify holes and species 2 to get a one-species TASEP
  model for which the phase diagram is given in table
  \eqref{phase_diag_1-TASEP}. The boundary conditions read
\begin{eqnarray}
\begin{array}{c c c}
&\text{left boundary}\hspace{1cm}&\text{right boundary}\\[1ex]
(P_1): & (0,2) \xrightarrow{\hspace{2mm}\alpha\hspace{2mm}} 1\hspace{1cm} & 1 \xrightarrow{\hspace{2mm}1\hspace{2mm}} (0,2)\\[1ex]
(P_2): & (0,2) \xrightarrow{\hspace{2mm}\alpha\hspace{2mm}} 1\hspace{1cm} & 1 \xrightarrow{\hspace{2mm}\beta\hspace{2mm}} (0,2)
 \end{array}
\end{eqnarray}
\item One can identify species 1 and 2 to get another version of the
  one-species TASEP model. In that case the two models $(P_1)$
  and $(P_2)$ produce the same boundary conditions:
\begin{eqnarray}\label{identif}
\begin{array}{c c c}
&\text{left boundary}\hspace{1cm}&\text{right boundary}\\[1ex]
(P_1)\,\&\,
(P_2): & 0 \xrightarrow{\hspace{2mm}1\hspace{2mm}} (1,2)\hspace{1cm} & (1,2) \xrightarrow{\hspace{2mm}\beta\hspace{2mm}} 0
 \end{array}
\end{eqnarray}
\end{enumerate}

 For the two-species TASEP, we denote by $J_1$, $J_2$
the particle currents  in the stationary state for the particles of
species 1 and 2 respectively.  The currents are counted positively
when particles flow from the left to the right.  In the same way, $\rho_1$ and $\rho_2$ denote the densities of particles of
species 1 and 2 respectively. 

\begin{figure}[htb]
\begin{center}
\subfloat[Boundary conditions  $(P_1)$]{
 \begin{tikzpicture}[scale=2/3]
\draw[->,very thick] (0,0)--(10,0);
\draw[->,very thick] (0,0)--(0,10);
\draw (5,0)--(5,10);
\draw (0,5)--(10,5);
\draw[ dotted, thick] (0,0)--(5,5);
\node at (9,-0.5) [] {$\alpha$} ;
\node at (-0.5,9) [] {$\beta$} ;
\node at (5,-0.5) [] {$\frac{1}{2}$} ;
\node at (-0.5,5) [] {$\frac{1}{2}$} ;
\node at (1,9) [] {\LARGE{IV}};
\node at (6,9) [] {\LARGE{I}};
\node at (1,4) [] {\LARGE{III}};
\node at (6,4) [] {\LARGE{II}};
\node at (2.5,7.5) [] {$J_1=\alpha(1-\alpha)$};
\node at (2.5,6.5) [] {$J_2=(\alpha-\frac12)^2$};
\node at (7,7.5) [] {$J_1=\frac14$};
\node at (7,6.5) [] {$J_2=0$};
\node at (2.5,2.5) [] {$J_1=\alpha(1-\alpha)$};
\node at (2.5,1.5) [] {$J_2=J_2^{(III)}$};
\node at (7,2.5) [] {$J_1=\frac14$};
\node at (7.4,1.5) [] {$J_2=-(\beta-\frac12)^2$};
 \end{tikzpicture}
}\hfill
\subfloat[Boundary conditions $(P_2)$]{
 \begin{tikzpicture}[scale=2/3]
\draw[->,very thick] (0,0)--(10,0);
\draw[->,very thick] (0,0)--(0,10);
\draw (5,5)--(5,10);
\draw (0,5)--(10,5);
\draw (0,0)--(5,5);
\draw (5,0)--(5,0.4);
\node at (9,-0.5) [] {$\alpha$} ;
\node at (-0.5,9) [] {$\beta$} ;
\node at (5,-0.5) [] {$\frac{1}{2}$} ;
\node at (-0.5,5) [] {$\frac{1}{2}$} ;
\node at (1,9) [] {\LARGE{IV}};
\node at (6,9) [] {\LARGE{I}};
\node at (1,2) [] {\LARGE{III}};
\node at (6,4) [] {\LARGE{II}};
\node at (2.5,7.5) [] {$J_1=\alpha(1-\alpha)$};
\node at (2.5,6.5) [] {$J_2=(\alpha-\frac12)^2$};
\node at (7,7.5) [] {$J_1=\frac14$};
\node at (7,6.5) [] {$J_2=0$};
\node at (2.2,4.4) [] {$J_1=\alpha(1-\alpha)$};
\node at (1.7,3.5) [] {$J_2=J_2^{(III)}$};
\node at (7,2.5) [] {$J_1=\beta(1-\beta)$};
\node at (7,1.5) [] {$J_2=0$};
 \end{tikzpicture}
}
\caption{Phase diagrams of the 2-TASEP with open boundaries for the boundary conditions
 $(P_1)$  and  $(P_2)$ (we have  used the notation $J_2^{(III)}=(\beta-\alpha)(1-\alpha-\beta)$).\label{fig:phase}}
 \end{center}
\end{figure}
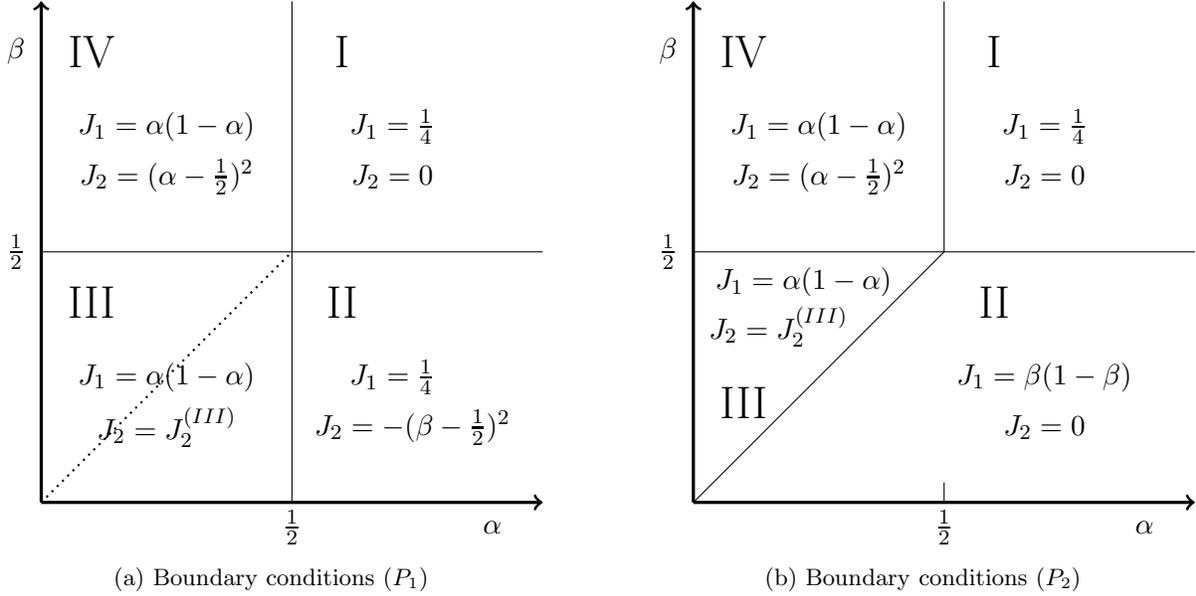

Identification 1. allows us to compute the current $J_1$ and the
density $\rho_1$, while identification 2. yields the current
$J_1+J_2\equiv-J_0$ and the density $\rho_1+\rho_2$. Gathering these
results, we obtain the phase diagrams depicted in figure
\ref{fig:phase}.

\subsubsection{Phase diagram of the  $(P_1)$ model}

 The phase diagram of the  $(P_1)$ model is displayed  in
 Figure~\ref{fig:phase}(a).  It comprises   four phases.   Using the
 identification procedure, we observe that  $\rho_1$ behaves as the
 density for the one-species TASEP with boundary rates $(\alpha,1)$,
 while $\rho_1+\rho_2$ behaves as the density for the one-species
 TASEP with boundary rates $(1,\beta)$.  The values of the  currents
 in each phase (see fig~\ref{fig:phase}(a)) are readily  found by this
 identification. The behaviour of the density profile ($\rho(j)$) in each phase
 can be  found in \cite{BE}.
\begin{description}
\item[Phase I: ]  For $\alpha>\frac{1}{2}$ and $\beta>\frac{1}{2}$, 
 first-class particles exhibit a maximal current,
  whereas the current of the second-class particles vanishes. The bulk
 density of  first-class  particles and  of holes is equal to 1/2, while
 the number of second class particles in the bulk is vanishingly small.
The density profiles of first and second class particles are characterised by
power law decays to the bulk values:
\beq
\rho_1=\frac12+\frac{1}{2\sqrt{\pi j}} +\cO\left(\frac1{j^{3/2}}\right) \mb{and} \rho_2=\cO\left(\frac1{j^{3/2}}\right),
\eeq
where $j$ is the site position on the lattice.
 The system is similar to the   one-species TASEP in its maximal current
 phase.
\item[Phase II:]  For $\alpha>\frac{1}{2}$ and $\beta<\frac{1}{2}$, 
  none of the currents $J_1,J_2$ and $J_0$ vanishes. This is a genuine
 2-TASEP phase with  boundaries  permeable to all the species. 
 The  two species  and the holes coexist in the bulk
 with non-zero bulk densities and density profiles characterised by power-law decays:
\beq
\rho_1=\frac12+\frac{1}{2\sqrt{\pi j}} +\cO\left(\frac1{j^{3/2}}\right) \mb{and} \rho_2=\frac12-\beta-\frac{1}{2\sqrt{\pi j}}+\cO\left(\frac1{j^{3/2}}\right).
\eeq
 First-class  particles are in their maximal
 current phase.
 Boundary effects are long-range for species 1 and 2.
\item[Phase III:] For $\alpha<\frac{1}{2}$ and $\beta<\frac{1}{2}$, 
 we obtain a `massive' phase in which  boundary effects are localized: 
 after a finite correlation length,   the system reaches  its bulk behaviour,
 \beq
\rho_1=\alpha+\cO\left(\frac1{j^{3/2}}\,\exp\left(-\frac{j}{\xi}\right)\right)  \mb{and} 
\rho_2=1-\alpha-\beta+\cO\left(\frac1{j^{3/2}}\,\exp\left(-\frac{j}{\xi}\right)\right).
\eeq
 The current of second-class particles $J_2$ vanishes along the line
 $\alpha = \beta <\frac{1}{2}$  and changes its sign across this line.
\item[Phase IV:] This phase, obtained for  $\alpha<\frac{1}{2}$
  and $\beta>\frac{1}{2}$,  is massive  for first-class  particles
 but  `massless' (exhibiting long-range correlations characterised by power laws) for  second-class particles and holes. Here again, the
  two species  and the holes  coexist in the bulk:
 \beq
\rho_1=\alpha+\cO\left(\frac1{j^{3/2}}\,\exp\left(-\frac{j}{\xi}\right)\right)  \mb{and} 
\rho_2=\frac12-\alpha-\frac{1}{2\sqrt{\pi j}}+\cO\left(\frac1{j^{3/2}}\right).
\eeq
Holes are in their  maximal current phase $J_0 = -1/4$.

\end{description}

\subsubsection{Phase diagram of the  $(P_2)$ model}

The phase diagram of the  $(P_2)$ model also  comprises   four phases,
displayed in Figure~\ref{fig:phase}(b). The diagram is qualitatively
different from that of the  $(P_1)$ model.  Here,  $\rho_1$ behaves as
the density for the one-species TASEP with boundary rates
$(\alpha,\beta)$, while $\rho_1+\rho_2$ behaves as the density for the
one-species TASEP with boundary rates $(1,\beta)$.
\begin{description}
\item[Phase I:]  For $\alpha>\frac{1}{2}$ and $\beta>\frac{1}{2}$, 
 first-class particles exhibit a maximal current. This phase is
 similar  to Phase I of model $(P_1)$.
\item[Phase II:] This phase is obtained for  $\beta <\alpha <\frac{1}{2}$. 
First-class particles are in their high density phase. The  bulk density of
 second-class particles in the bulk vanishes;  moreover, the
 probability to find a second class particle at a distance larger
 than  the correlation length $\xi$  away from the  boundaries, 
 is exponentially  small: 
\beq
\rho_1=1-\beta+\cO\left(\frac1{j^{3/2}}\,\exp\left(-\frac{j}{\xi}\right)\right)  \mb{and} 
\rho_2=\cO\left(\frac1{j^{3/2}}\,\exp\left(-\frac{j}{\xi}\right)\right) .
\eeq
\item[Phase III:]  For  $\alpha < \beta <\frac{1}{2}$, the two species and the holes
 are simultaneously present with non-vanishing currents. 
 The current of second-class particles $J_2$ is strictly positive.
 This phase is  massive for the two  classes of particles and the holes: 
\beq
\rho_1=\alpha+\cO\left(\exp\left(-\frac{j}{\xi}\right)\right)  \mb{and} 
\rho_2=1-\alpha-\beta+\cO\left(\,\exp\left(-\frac{j}{\xi}\right)\right) .
\eeq
\item[Shock Line:] This line corresponds to  $\alpha = \beta <\frac{1}{2}$.
The density profiles $\rho_1$ and $\rho_2$ display a linear behaviour
that reflect a coexistence  between a  low density and a high density
 regions: 
  \beq
\rho_1= \alpha+\frac{j}{L}(1-2\alpha)  \mb{and} 
\rho_2=(1-\alpha)(1-\frac{j}{L}).
\eeq
The density profile of  first-class particles takes the values
 $\alpha$ and  $1-\alpha$ with a discontinuous shock between the two
 regions. The second-class particles have a plateau density of
 $1-\alpha$ to the left of the shock  and zero  density  to the right shock. 
This means effectively
 that in the stationary state in the infinite system limit only the left reservoir is active
 as far as second-class particles  are concerned.
\item[Phase IV] This phase, obtained for  $\alpha<\frac{1}{2}$
  and $\beta>\frac{1}{2}$, is similar to   Phase IV
 of the  $(P_1)$  model.
\end{description}

\section{Summary of  matrix product solutions}

 In this section,  the  stationary state of the 2-TASEP with open
 boundaries  is represented as a   matrix state, for the models
 $(P_1)$ and  $(P_2)$. We show that  the  steady-state probability of
 finding a given configuration $(\sigma_1,\sigma_2,...,\sigma_L)$ can
 be  written as  a contraction on two vectors $\llangle W|$ and
 $|V\rrangle$ over a suitable algebra 
 \beq
 P(\sigma_1,\sigma_2,...,\sigma_L) = \frac1{Z_L} \llangle W|
 X_{\sigma_1}\,X_{\sigma_2}\,\cdots X_{\sigma_L}\,|V\rrangle\,,
\label{AnsMat}
\eeq
  where $Z_L$ is a normalisation constant ensuring 
 that $P(\sigma_1,\sigma_2,...,\sigma_L)$ is a probability.

 In the following,  we give  explicit formulas for the 
 operators   $X_{\sigma}$ that generate the algebra 
 and  for the boundary vectors $\llangle W|$ and 
 $|V\rrangle$. The operators   $X_{\sigma}$  and the left vector 
 $\llangle W|$ are the same for both models  $(P_1)$ and  $(P_2)$;
 only the right vectors  $|V\rrangle$ differ. We shall also
 present an important factorisation property of the Matrix Ansatz
  that will allow us to derive  explicit expressions for
 the normalisation constant  $Z_L$.  The proofs
 of these results will be given in section \ref{sect:proof}.

\subsection{Explicit representation  of  the matrices}

 The Matrix Ansatz for the 2-TASEP will be constructed  in terms of
 tensor products of the fundamental operators $A$, $\delta$ and
 $\varepsilon$ that appear in the solution of the one-species
 exclusion process \cite{DEHP}.  These operators  $A$, $\delta$ and
 $\varepsilon$ define a quadratic algebra and satisfy  
 \beq
 \delta\,\varepsilon = 1\,,\quad A=1-\varepsilon\,\delta\,,\quad
 \delta\,A = 0\,,\quad  A\,\varepsilon = 0\,.\label{eq:ed} 
 \eeq 
 The
 relation with  the operators $D$ and $E$ of  \cite{DEHP}  is  $\delta
 = D -1, \,  \varepsilon = E -1 $ and $ A = DE - ED$. 

\hfill\break
 We also define  the following parameters 
\begin{equation}
a=\frac{1-\alpha}{\alpha}\quad \mbox{and}\quad  b=\frac{1-\beta}{\beta}\;.
\end{equation}

\hfill\break
Finally, we shall   need four  commuting copies of the algebra \eqref{eq:ed}, $(\varepsilon_n,\delta_n,A_n)$, $n=1,2,3,4$.
A simple way to achieve this  is to make  four-fold
tensor products:
\beq
\begin{aligned}
\varepsilon_1 = \varepsilon \otimes 1\otimes 1 \otimes 1,  \\ 
\varepsilon_2 = 1 \otimes \varepsilon\otimes 1 \otimes 1 ,  \\ 
\varepsilon_3 = 1\otimes 1 \otimes\varepsilon \otimes  1,  \\ 
\varepsilon_4 = 1\otimes 1\otimes 1 \otimes\varepsilon ,
\label{tensor2}
\end{aligned}
\eeq
and similarly for $\delta_n$ and $A_n$.

\hfill\break

 We are now in a position to present explicit matrices for the 2-TASEP
 with open boundaries
\vskip 0.2cm 
 \begin{eqnarray}
 X_0 &=&\Big(1+a A_1 A_2+\varepsilon_2 \delta_3 \Big)(1+\varepsilon_4) + \Big( \varepsilon_2+\varepsilon_3+a A_1A_2\varepsilon_3+\varepsilon_1A_3\Big)(1+\delta_4),
\label{X0} 
\\
X_2 &=&a \delta_1A_2\,(1+\varepsilon_4)+ \Big(  a \delta_1A_2\varepsilon_3+a A_2A_3\Big)(1+\delta_4),
\\
X_1 &=&\Big( \delta_2+\delta_3 \Big)(1+\varepsilon_4)+ \Big( 1+\delta_2\varepsilon_3+\varepsilon_1\delta_2A_3 \Big)(1+\delta_4).
\label{X2}
\end{eqnarray}
\vskip 0.2cm 

\noindent 
{\bf N. B.} It is important to realise that the integer suffices on
the left and right hand sides of  \eqref{X0}--\eqref{X2}  are unrelated:
on the left $\sigma =0, 1, 2$  corresponds to particle species
whereas on the right $n=1, 2, 3, 4$  labels the tensor product as in
the example (\ref{tensor2}).

\subsection{Expressions of the boundary vectors}

 To construct the vectors $\llangle W|$ and $|V\rrangle$, we first  define the elementary vectors
$\llangle x|$ and $|x\rrangle$ that obey
\beq
\llangle x|\,\varepsilon = x\,\llangle x| \mb{and} \delta\,|x\rrangle = x\,|x\rrangle\,.
\label{eq:vectx}
\eeq
It is known \cite{DEHP} that explicit representations of such elementary vectors exit. Here we use a representation where 
\beq
\llangle x|y\rrangle = \frac1{1-xy}\,.
\eeq
The left boundary  vector reads:
\begin{equation}
  \llangle W|_{1234} =\llangle 1|_1\llangle 0|_2\llangle 0 |_3\llangle 0|_4\;,
\end{equation}
where the indices indicate again which copy of the $(A,\delta,\eps)$
algebra acts on the vector.  To make the notation less cluttered, we
shall simply write  $\llangle W|$  instead of  $\llangle
W|_{1234}$. Note that the left  vector is the same  for  the models
$(P_1)$ and  $(P_2)$.

  The right boundary  vector  depends on the choice of the dynamics at
  the  right boundary  (i.e. on the choice of the model $(P_1)$ or
  $(P_2)$). We have 
 \begin{eqnarray}
 |V (P_1) \rrangle_{1234} &=& |\frac{b}{a}\rrangle_1\, |0\rrangle_2\,
|1\rrangle_3\, |b\rrangle_4 \,\, \,\mb{for} (P_1) \\ 
\mb{and} |V (P_2)\rrangle_{1234} &=& \, |0\rrangle_1\, |b\rrangle_2 \,|1\rrangle_3\,
|b\rrangle_4  \,\,\,\,  \mb{for} (P_2).
\end{eqnarray}
 We shall simply   write  $|V\rrangle$ for the right vector, without
 specifying  the  indices and  which model we consider. This should
 be unambiguous from the context.

\subsection{A factorisation property of the matrix ansatz \label{FactLL2-3}} 

The expressions \eqref{X0}--\eqref{X2} for the $X_\sigma$'s can be written in a factorized form
which will be useful to compute the normalisation and for the proof of the
matrix ansatz in section \ref{sect:proof}.

 Let us consider the following  matrices \beq
\begin{array}{lll}
&\LL^{(3)} = \begin{pmatrix}
1+\lambda A_1A_2+\varepsilon_2\delta_3 & \varepsilon_2+\varepsilon_3+\lambda A_1A_2\varepsilon_3+\varepsilon_1A_3 \\
\lambda \delta_1A_2 & \lambda \delta_1A_2\varepsilon_3+\lambda A_2A_3\\
\delta_2+\delta_3 & 1+\delta_2\varepsilon_3+\varepsilon_1\delta_2A_3
\end{pmatrix}
\,,
&\LL^{(2)}=\begin{pmatrix}
1+\varepsilon_4\\ 1+\delta_4
\end{pmatrix},
\label{def:LL2-LL3}
\end{array}
\eeq 
where $\LL^{(3)}$ is a $3\times 2$ matrix which contains a parameter $\lambda$, while $\LL^{(2)}$ is a
$2\times 1$ matrix.  Then, for $\lambda=a$,  the following,
important,  relation is satisfied 
\beq \label{facto1}
\begin{pmatrix} X_{0} \\ X_{2} \\ X_{1}\end{pmatrix} = \LL^{(3)}\,\LL^{(2)}
\eeq
 where $X_0$, $X_2$ and $X_1$ are the operators that perform the
 matrix Ansatz for the open 2-TASEP. This identity can readily be
checked using equations  \eqref{X0}--\eqref{X2}.

Furthermore, the operator  $\LL^{(3)}$, defined above,
can be  factorized into the product of a $3\times3$ matrix 
by a $3\times 2$ matrix, as follows
\beq  \label{factoriLL3}
\LL^{(3)}=L^{(3)}\,\wt L^{(3)}
\eeq
 with
\begin{equation} \label{factoriLL3expli}
 L^{(3)}=\begin{pmatrix}
     1+ \lambda A_1 A_2& \varepsilon_1& \varepsilon_2
\\  \lambda \delta_1  A_2& \lambda A_2&0
\\  \delta_2& \varepsilon_1 \delta_2&1
   \end{pmatrix}\mb{and}  \wt L^{(3)}=\begin{pmatrix}
     1&\varepsilon_3\\ 
    0&A_3\\
    \delta_3&1
   \end{pmatrix}.
\end{equation}
A similar  type of factorisation   holds  for $\LL^{(2)}$:
\begin{equation} \label{factoriLL2}
 \LL^{(2)}= L^{(2)}\wt L^{(2)}
\end{equation}
with
\begin{equation}\label{factoriLL2expli}
 L^{(2)}=\begin{pmatrix}
            1 & \varepsilon_4\\
            \delta_4& 1
           \end{pmatrix}
           \mb{and}
\wt L^{(2)}=  \begin{pmatrix}
            1 \\
            1
           \end{pmatrix}.
\end{equation}
The origin of these factorisations will be clarified in 
Section \ref{sect:proof}.  Note that the factorisations are valid for
arbitrary values of the parameter $\lambda$. We present also in Appendix \ref{app:LLt} 
a relation between $\wt L$ and $L$. 

\subsection{Factorisation property of the steady-state probabilities} 
  A compact way of writing the
 steady-state probabilities is to define a vector~:
\beq\label{eq:XX}
\boldsymbol{X}= \begin{pmatrix} X_{0} \\ X_{2} \\ X_{1}\end{pmatrix}\, .
\eeq
Then, the matrix Ansatz for the stationary probability vector reads
\beq
 \bP = \frac{1}{Z_L} 
 \llangle W|\boldsymbol{X} \otimes \boldsymbol{X}\otimes 
 \cdots\otimes \boldsymbol{X}\,|V\rrangle\,.
 \label{eq:M3}
\eeq
 The factorisation \eqref{facto1} leads to 
\beq \label{eq:Pfac}
\bP = \frac1{Z_L} \PP^{(3)}\    \PP^{(2)} \,,
\eeq
where
\begin{equation}
 \PP^{(3)}=
 \llangle W|_{123}\ \LL^{(3)}\otimes \cdots \otimes \LL^{(3)}\,|V\rrangle_{123}\mb{and}
 \PP^{(2)}=
 \llangle W|_{4}\ \LL^{(2)}\otimes\cdots\otimes \LL^{(2)}\,|V\rrangle_{4}\;.
\end{equation}
 Here,  $\PP^{(3)}$ is a $3^L \times 2^L$ matrix and $\PP^{(2)}$
  is a $2^L$-component vector so that $\bP$ is a $3^L$-component
  vector as expected. We also remark that $\PP^{(2)}$ (up to a normalisation)
 is identical to the steady-state vector of the one species TASEP
  with open boundaries. Therefore,  we have 
\begin{equation} \label{Stat1TASEP}
  M^{(2)} \PP^{(2)}=0 \;,
\end{equation}
where $M^{(2)}$ is the Markov matrix of the one-species TASEP.

 Finally,   thanks to the  factorisation properties of 
\eqref{factoriLL3}--\eqref{factoriLL2expli} of $\LL^{(3)}$ and $ \LL^{(2)}$,
  the stationary state can   be further  decomposed as 
\begin{eqnarray}
\bP &=&  \frac1{Z_L} \PP^{(3)}\    \PP^{(2)}=\frac1{Z_L}  P^{(3)}\ \wt P^{(3)}\ P^{(2)}\ \wt P^{(2)}\;,
\end{eqnarray}
with 
\begin{eqnarray}
 &&P^{(3)}=\llangle W|_{12}\ L^{(3)}\otimes L^{(3)} \otimes\cdots \otimes L^{(3)}\,|V\rrangle_{12} \\
 &&\wt P^{(3)}=\llangle W|_{3}\ \wt L^{(3)}\otimes \wt L^{(3)}\otimes\cdots \otimes \wt L^{(3)}\,|V\rrangle_{3}\\
 &&P^{(2)}=\llangle W|_{4}\ L^{(2)}\otimes L^{(2)}\otimes\cdots \otimes L^{(2)}\,|V\rrangle_{4} \\
 &&\wt P^{(2)}=\begin{pmatrix}
            1 \\
             1
           \end{pmatrix}\otimes \begin{pmatrix}
            1 \\
             1
           \end{pmatrix}\otimes\dots \otimes \begin{pmatrix}
            1 \\
             1
           \end{pmatrix} \;.
\end{eqnarray}
Let us note that $P^{(3)}$ is a $3^L \times 3^L$ matrix, $\wt P^{(3)}$ is a $3^L \times 2^L$ matrix, $P^{(2)}$ is a $2^L\times 2^L$ matrix and
$\wt P^{(2)}$ is a $2^L$-component vector with constant components.

\subsection{Calculation of the normalisation \label{sect:part-fct}}
We may now use the factorisation 
properties  of the previous subsection to calculate the normalisation $Z_L$ of the stationary
probabilities  \eqref{AnsMat}. 
The results we  obtain  are
\beq \label{eq:ZL}
 \begin{aligned}
 &Z_L = \frac{a}{a-b}\,\cZ_L(\alpha,1)\,\cZ_L(1,\beta)  \,\,\,\,  \mb{ for } (P_1)
 \\[1ex]
 &Z_L = (1-ab)\,\cZ_L(\alpha,\beta)\,\cZ_L(1,\beta) \mb{ for } (P_2)
 \end{aligned}
 \eeq where $\cZ_L(\alpha,\beta)$ is the partition function of the
 open one-species TASEP with injection rate $\alpha$ and extraction
 rate $\beta$. Its exact  expression \cite{BE} is given by
 \beq
  \label{Norme1TASEP}
 \cZ_L(\alpha,\beta) =  \llangle
 a|(2+\varepsilon+\delta)^L|b\rrangle =  \llangle a|(D +
 E)^L|b\rrangle =  \sum_{p=0}^L \frac{ p \, (2 L - p -1)!}{L! (L-p)!}
 \frac{ \left( \frac{1}{\alpha} \right)^{p+1} -  \left(
   \frac{1}{\beta} \right)^{p+1} } { \frac{1}{\alpha}  -
   \frac{1}{\beta}  } \,\llangle a|b\rrangle \,  .  \eeq 
 From the matrix Ansatz, we know that   
 \beq 
 Z_L=\llangle
 W|\,(X_0+X_1+X_2)^L\,|V\rrangle.  
 \eeq 
 Using  the factorisations
 \eqref{facto1} and \eqref{factoriLL3}, we obtain  \beq X_0+X_1+X_2=
 (1,1,1)\cdot \begin{pmatrix} X_{0} \\ X_{2} \\ X_{1}\end{pmatrix} \,
  = (1,1,1) L^{(3)}\,\wt L^{(3)} \LL^{(2)} . 
  \eeq 
  We  first  compute
 \beq (1,1,1)\cdot L^{(3)} =
 \Big(1+a(A_1+\delta_1)A_2+\delta_2\,,\ \varepsilon_1(1+\delta_2)
 +a A_2,\ 1+\varepsilon_2\Big).
 \eeq 
 Then, from  the relations $\llangle 1|(A+\delta)=\llangle 1|$
 and $\llangle 1|\varepsilon=\llangle 1|$,  we deduce
 \beq
 \llangle  1|_1 (1,1,1)\cdot L^{(3)} =
 \Big(1+aA_2+\delta_2\,,\ 1+aA_2+\delta_2,\ 1+\varepsilon_2\Big)\llangle
 1|_1
 \eeq 
 This implies  that the space 1 drops out (because neither
 $\wt L^{(3)}$ nor $ \LL^{(2)}$ act on it).  Remarking that  
 \bea
 \Big(1+aA_2+\delta_2\,,\ 1+aA_2+\delta_2,\ 1+\varepsilon_2\Big)\wt
 L^{(3)} &=&
  \Big(1+aA_2+\delta_2\,,\ 1+\varepsilon_2\Big)\, 
\begin{pmatrix} 1 &1 & 0 \\ 0&0&1\end{pmatrix} 
 \wt L^{(3)} \qquad
\nonu
 &=& \Big(1+aA_2+\delta_2\,,\ 1+\varepsilon_2\Big)\, 
 \begin{pmatrix} 1 &A_3+\varepsilon_3 \\ \delta_3&1\end{pmatrix}
\eea
  and using $(A+\varepsilon)|1\rrangle=|1\rrangle$
  and $\delta|1\rrangle=|1\rrangle$, we have 
\beq
 \begin{pmatrix} 1 &A_3+\varepsilon_3 \\ \delta_3&1\end{pmatrix}|1\rrangle_3 
 = |1\rrangle_3 \begin{pmatrix} 1 &1 \\ 1&1\end{pmatrix}
 =|1\rrangle_3 \begin{pmatrix} 1  \\ 1\end{pmatrix} ( 1 ,1 )
 \eeq
 so that space 3 also drops out. Gathering the different results, we obtain 
 \beq
 \begin{cases}
 Z_L = \llangle 1|b/a\rrangle_1\  \llangle 0|(2+aA_2+\delta_2+\varepsilon_2)^L
 |0\rrangle_2 \ \llangle 0 |1\rrangle_3\
 \llangle 0|(2+\varepsilon_4+\delta_4)^L|b\rrangle_4
  \quad \mbox{ for } (P_1)
 \\[1ex]
 Z_L = \llangle 1|0\rrangle_1\  \llangle 0|(2+aA_2+\delta_2+\varepsilon_2)^L|b\rrangle_2 \ \llangle 0 |1\rrangle_3\
 \llangle 0|(2+\varepsilon_4+\delta_4)^L|b\rrangle_4
   \quad \quad \, 
  \mbox{ for }  (P_2). \end{cases}
 \eeq
  We conclude the derivation  of \eqref{eq:ZL} by using  \eqref{Norme1TASEP} and by observing that 
\beq\label{norm.ZL}
 \llangle 0|(2+aA+\delta+\varepsilon)^L|b\rrangle=
 \frac{\llangle 0|b\rrangle}{\llangle a|b\rrangle}\cZ_L(\alpha,\beta) 
\eeq
because  the operators $\wt\eps=aA+\eps$ and $\delta$ obeys 
the same algebraic rules as $\eps$ and $\delta$, but
now  $\llangle 0|$ is a left eigenvector of   $\wt\eps$ with eigenvalue $a$. 
  
\section{Proof of the matrix ansatz\label{sect:proof}}

 In this section, we give an algebraic proof that the matrix Ansatz
 given in the previous section is indeed a representation of the
 steady-state probabilities of the models $(P_1)$ and $(P_2)$.  We
 shall use the method of auxiliary matrices \cite{BE,sandow,nikolaus}
 that set out a general cancellation scheme and led Krebs and Sandow
 \cite{KS} to  a general proof (albeit not constructive) of the
 matrix-product form  for a general class of stochastic processes. 

\subsection{Local update operators}

 The evolution  rules of the exclusion process are local:
 a particle moves to one of its neighbouring sites.  Hence,  the Markov matrix 
 of the process can be  written as the sum of local operators. For the
  one-species TASEP  with open boundaries, we have 
\begin{equation}
M^{(2)} =  B_1^{(2)} + \sum_{\ell=1}^{L-1} m^{(2)}_{\ell,\ell+1} + \overline B_L^{(2)}\,,
\label{Markov1species}
\end{equation}
 where  the  local bulk  Markov matrix between site $\ell$ and $\ell+1$
 and the  boundary matrices  are given by\footnote{
We present here only the boundary matrix $B^{(2)}$ with an injection rate $\alpha=1$ that is needed for our purposes, see relations \eqref{eq:b1} and \eqref{eq:b1bis}. Remark that the matrices $B^{(2)}$ and $\overline B^{(2)}$ correspond to identification \eqref{identif}.}
\begin{equation}\label{markov2}
 m^{(2)}=\begin{pmatrix}
          0&0&0&0\\
          0&0&1&0\\
          0&0&-1&0\\
          0&0&0&0
         \end{pmatrix}\ ;\quad  B^{(2)}=\begin{pmatrix} -1& 0 \\ 1& 0 \end{pmatrix}\ ;\quad 
         \overline B^{(2)}=\begin{pmatrix} 0& \beta \\ 0& -\beta \end{pmatrix}\;.
\end{equation}

Similarly, the dynamics of the 2-species TASEP 
 is governed by the Markov matrix $M^{(3)}$. It can be decomposed as 
\begin{equation}
\label{decomM3}
M^{(3)} =  B_1 + \sum_{\ell=1}^{L-1} m^{(3)}_{\ell,\ell+1} + \overline B_L\,,
\end{equation}
with the local bulk update operator acting on nearest 
neighbour sites  
\begin{eqnarray}
m^{(3)}=\begin{pmatrix}
   .&.&.&.&.&.&.&.&.\\
   .&.&.&1&.&.&.&.&.\\
   .&.&.&.&.&.&1&.&.\\
   .&.&.&-1&.&.&.&.&.\\
   .&.&.&.&.&.&.&.&.\\
   .&.&.&.&.&.&.&1&.\\
   .&.&.&.&.&.&-1&.&.\\
   .&.&.&.&.&.&.&-1&.\\
   .&.&.&.&.&.&.&.&.
  \end{pmatrix}
\end{eqnarray}
where the points in the matrix stand for vanishing entries. The boundary operators read
\begin{eqnarray}\label{eq:defB}
 B=\begin{pmatrix}
 -1   &0 &0\\
 1-\alpha & -\alpha &0\\
 \alpha& \alpha&0
    \end{pmatrix}\,,
 \qquad   \wh B=\begin{pmatrix}
 0  &\beta &\beta\\
 0 & -\beta &1-\beta\\
 0& 0&-1
    \end{pmatrix}\,,\qquad    
 \wt  B=\begin{pmatrix}
 0  &\beta &\beta\\
 0 & -\beta &0\\
 0& 0&-\beta
    \end{pmatrix}\,.
\end{eqnarray}
These operators are written in the local state basis $(0,2,1)$ which is 
 the  natural choice corresponding  to increasing order of priority
 in the update rules.
In equation \eqref{decomM3}, the 
 subscripts indicate on which sites of the lattice
 the local operators act non-trivially,
and the right boundary  matrix $\overline B$ corresponds to
   $\wh B$  for the processes $(P_1)$  and  $\wt B$ for $(P_2)$. 
As a rule, the superscripts in \eqref{markov2} and
  \eqref{decomM3} indicate
 the number of possible states at a site,  i.e. the number of species plus one.
 However,  to lighten the notation we do not put a superscript $(3)$
  in the boundary matrices for the 2-TASEP, defined in  \eqref{eq:defB}.


\subsection{Auxiliary matrices}

 We want to prove that  \eqref{eq:M3} is a representation
 of  the  stationary vector
 of the 2-TASEP with open boundaries, i.e.  that 
 the master equation \eqref{master_equation} is satisfied. 
 The  matrix Ansatz  has a 
 straightforward algebraic proof \cite{sandow,KS,nikolaus}: if  
 one can find auxiliary operators 
 $\boldsymbol{X'}= \begin{pmatrix} X'_{0} \\ X'_{2} \\ X'_{1}\end{pmatrix}$ 
 such that  $\boldsymbol{X}$ and $\boldsymbol{X'}$  satisfy
\bea\label{eq:mXX}
&&m^{(3)}\,\boldsymbol{X}\otimes\boldsymbol{X} = 
\boldsymbol{X'}\otimes \boldsymbol{ X}-\boldsymbol{X}\otimes\boldsymbol{X'}  \\
\label{eq:BXX}
&& \llangle W| B\,\boldsymbol{X}= -\llangle W|\boldsymbol{X'}
\mb{and} \overline B \boldsymbol{X}|V\rrangle = \boldsymbol{X'}|V\rrangle
\eea
 where we recall that $\overline B$ is  either  $\wh B$ or $\wt B$, then 
 the stationary master equation \eqref{master_equation} is satisfied for 
 for the stationary probability vector given by \eqref{eq:M3}.
 
 Before giving an explicit realisation of these new operators $X'_0$, $X'_1$ and $X'_2$, 
 we want to explain the notations. 
 The auxiliary generators $X'_{0}, X'_{2}$ and  $X'_{1}$ are often  denoted
 by a hat or a bar in the literature (and nicknamed `hat-matrices'). However,
 we have purposely written them with a prime, because  we shall show 
 in  section \ref{sec:ZF} that  the  $\boldsymbol{X'}$ can be constructed by 
 taking the  derivative with respect to a spectral parameter
 of the Zamolodchikov-Faddeev relation. 
 Recalling that the tensor product
 in equation \eqref{eq:mXX} is given by 
\beq
\boldsymbol{X}{\otimes} \boldsymbol{X'} = \begin{pmatrix}
X_0X'_0 \\
X_0X'_2 \\
X_0X'_1 \\
X_2X'_0 \\
X_2X'_2 \\
X_2X'_1 \\
X_1X'_0 \\
X_1X'_2 \\
X_1X'_1
\end{pmatrix}\,, 
\eeq
 we can spell out the formula \eqref{eq:mXX} to  obtain 
 the quadratic  relations that couple 
 $X_{0}, X_{2}$ and  $X_{1}$  with the auxiliary matrices 
 $X'_{0}, X'_{2}$ and  $X'_{1}$.
 \bea
 && [X_i\,,\,  X'_i]=0\,,\quad i=0,1,2\\
 && X_1\,X_0 =X'_0 X_1 -  X_0X'_1   =X_1X'_0- X'_1 X_0\,,  \\
 && X_2\,X_0 =X'_0 X_2 -  X_0X'_2   =X_2X'_0- X'_2 X_0\,, \\
 && X_1\,X_2 =X'_2 X_1 -  X_2X'_1   =X_1X'_2- X'_1 X_2\,.
 \eea
In Appendix \ref{app:CMRV}, we give the connection between the bases presented here and the ones introduced in \cite{CMRV}.

\subsection{Explicit formulas for the auxiliary matrices}

As was done for the matrices $X_i$, we wish to express the  auxiliary matrices  $X'_i$ in terms
 of tensor products of the fundamental operators $A$, $\delta$ and $\varepsilon$ satisfying the defining relations \eqref{eq:ed}. We have seen,  
in \eqref{facto1}, that 
$\boldsymbol{X}=\LL^{(3)}\,\LL^{(2)}$.  
 A similar representation for  $\boldsymbol{X'}$ is given  by
\beq \label{eq:XXb}
\boldsymbol{X'}=\LL^{(3)'}\,\LL^{(2)}
+ \LL^{(3)}\,\LL^{(2)'}
\eeq
where  $\LL^{(3)}$ and $\LL^{(2)}$  have been defined in
  \eqref{def:LL2-LL3} and 
\beq \label{defiLL'}
\LL^{(3)'} = \begin{pmatrix}
1 & \varepsilon_1A_3+\varepsilon_3 \\
0 &0 \\
-\delta_3 & -1
\end{pmatrix}
 \quad \quad  \mb{and} \quad \quad
\LL^{(2)'} =
\begin{pmatrix}
1\\ -1
\end{pmatrix}.
\eeq
 As in \eqref{tensor2}, the generators $(\varepsilon_n,\delta_n,A_n)$, 
 $n=1,2,3,4$ are commuting copies of the algebra \eqref{eq:ed}.
At first sight, the sum expression \eqref{eq:XXb} for $\boldsymbol{X'}$  may 
 seems a bit arbitrary  but this form that is reminiscent of the derivative
 of a product will appear  natural in  section \ref{sec:ZF}.
Now,  using \eqref{facto1} and \eqref{eq:XXb}, 
one can verify by a direct  but lengthy calculation
 that the relations \eqref{eq:mXX} and \eqref{eq:BXX}  are  satisfied and 
 thus prove  the matrix Ansatz.  
 However, 
 the calculation can be simplified using 
 some  factorisations, as  explained  in the next subsection.


\subsection{Synthetic proof of the auxiliary algebra \label{sect:facto}}

  We have shown  in Section  \ref{FactLL2-3} that the matrices
 $\LL^{(3)}$ and $ \LL^{(2)}$ can be factorized, see equations  
 \eqref{factoriLL3}--\eqref{factoriLL2expli}. 
 We have a  corresponding  property that  holds for $ \LL^{(3)'}$: 
\begin{equation}\label{eq:LL3}
 \LL^{(3)'}=L^{(3)'}\wt L^{(3)}+L^{(3)}\wt L^{(3)'}
\end{equation}
with 
  \begin{equation}
 L^{(3)'}=\begin{pmatrix}
     1& \varepsilon_1& \varepsilon_2
\\  0& 0&0
\\  0& 0&0
   \end{pmatrix}
\mb{and}
\wt L^{(3)'}=\begin{pmatrix}
     0&0\\ 
    0&0\\
    -\delta_3&-1
   \end{pmatrix}
\end{equation}
and where $L^{(3)}$ and $\wt L^{(3)}$ are given  in 
 \eqref{factoriLL3expli}. 

Using the explicit form of these operators,  one can verify the following
 relations
\begin{eqnarray}
  m^{(3)} L^{(3)}\otimes L^{(3)}-L^{(3)}\otimes L^{(3)} m^{(3)}&=&  L^{(3)'}\otimes L^{(3)}- L^{(3)}\otimes L^{(3)'}\label{eq:mm1}\\
  m^{(3)} \wt L^{(3)}\otimes \wt L^{(3)}-\wt  L^{(3)}\otimes \wt  L^{(3)} m^{(2)}&=& \wt  L^{(3)'}\otimes \wt  L^{(3)}-\wt  L^{(3)}\otimes \wt L^{(3)'}\label{eq:mm2}
\end{eqnarray}
where $m^{(2)}$ is the local operator for the one-species TASEP,
 given in  \eqref{markov2}.  Recalling that 
 $\LL^{(3)}=L^{(3)}\wt L^{(3)}$ and 
$\LL^{(3)'}=L^{(3)'}\wt L^{(3)}+L^{(3)}\wt L^{(3)'}$, 
we conclude that 
\begin{eqnarray}\label{eq:tel3}
 m^{(3)} \LL^{(3)}\otimes \LL^{(3)}-\LL^{(3)}\otimes \LL^{(3)} m^{(2)}=  \LL^{(3)'}\otimes \LL^{(3)}- \LL^{(3)}\otimes \LL^{(3)'}\;.
\end{eqnarray}
 Similarly, from 
   the factorisation of  $\LL^{(2)}$,  \eqref{factoriLL2} and \eqref{factoriLL2expli},
 we can check that 
\begin{eqnarray}\label{eq:tel2}
 m^{(2)} \LL^{(2)}\otimes \LL^{(2)}=  \LL^{(2)'}\otimes \LL^{(2)}- \LL^{(2)}\otimes \LL^{(2)'}\;,
\end{eqnarray}
where  $\LL^{(2)'}$  has been defined in \eqref{defiLL'}.

 Combining 
equations \eqref{eq:tel3} and \eqref{eq:tel2} with expressions \eqref{eq:XXb}
 ends the proof of the bulk relation   \eqref{eq:mXX}.

 The following identities hold for $\LL^{(3)}$
\begin{eqnarray}\label{eq:b1}
&& \llangle W|_{123} \Big( B \LL^{(3)}-\LL^{(3)}  B^{(2)} \Big) =  -\llangle W|_{123}\LL^{(3)'}  \\
&& \label{eq:b1bis}
\Big( \overline B \LL^{(3)}- \LL^{(3)} \overline B^{(2)}\Big) |V\rrangle_{123}=  \LL^{(3)'}|V\rrangle_{123} \;.
\end{eqnarray}
Similar relations also exist for $\LL^{(2)}$
\begin{eqnarray}
 \llangle 0|_{4}\, B^{(2)} \LL^{(2)}=  -\llangle 0|_{4}\,\LL^{(2)'}  \\
 \overline B^{(2)} \LL^{(2)}|b\rrangle_{4}=  \LL^{(2)'} |b\rrangle_{4} \;.\label{eq:b2}
\end{eqnarray}
 Combining equations \eqref{eq:b1}-\eqref{eq:b2} leads us  to 
 the boundary relations \eqref{eq:BXX}.

 Finally, from the  identities \eqref{eq:tel3}, \eqref{eq:b1} and \eqref{eq:b1bis}, we deduce that 
 $M^{(3)} \PP^{(3)}=\PP^{(3)} M^{(2)}$, so that we obtain
\begin{equation}\label{eq:comMP}
 M^{(3)}\bP  =  M^{(3)} \PP^{(3)}   \PP^{(2)}  = \PP^{(3)} M^{(2)} \PP^{(2)} =0 \;,
\end{equation}
where we have used   \eqref{Stat1TASEP}.  
 This concludes the proof 
 that $\bP = \frac1{Z_L} \PP^{(3)}\    \PP^{(2)}$
 is the stationary state of the 2-TASEP with open boundaries.

\section{Zamolodchikov-Faddeev algebra and  Ghoshal-Zamolodchikov
 relation \label{sec:ZF}}


In this section, we show that the relations used in section \ref{sect:facto} can be obtained from a more general framework. 
The relations introduced 
in this section will depend on an additional parameter,
called the spectral parameter. The relations of the previous section will be recovered by setting this parameter to a specific value.

The main objects necessary in this section are similar to those used to prove the integrability of the Markov matrix $M^{(3)}$ 
(see \cite{sklyanin} for historical paper or \cite{CMRV} for the case treated here).
We need the R-matrix encoding the bulk dynamics and the K-matrices encoding the boundaries rates.

For the 2-species TASEP, the braided R-matrix reads $\check R^{(3)}(x)=1+(1-x)\, m^{(3)}$ with the property $-\check R^{(3)'}(1)=m^{(3)}$.
The K-matrix for the left boundary is
\begin{eqnarray}
  K(x)=\begin {pmatrix} {x}^{2}&0&0\\ \noalign{\medskip}-{\frac {a
x \left( {x}^{2}-1 \right) }{xa+1}}&{\frac {x \left( a+x \right) }{xa+
1}}&0\\ \noalign{\medskip}-{\frac {{x}^{2}-1}{xa+1}}&-{\frac {{x}^{2}-
1}{xa+1}}&1\end {pmatrix} 
\end{eqnarray}
and the ones for the two choices of right boundary are
\begin{eqnarray}
\wh K(x)=\begin {pmatrix}  1&{\frac {{x}^{2}-1}{ \left( b+x \right) x
}}&{\frac {{x}^{2}-1}{ \left( b+x \right) x}}\\ 
0&{\frac {bx+1}{ \left( b+x \right) x}}&{\frac {b \left( {x}^{2}-1
 \right) }{ \left( b+x \right) {x}^{2}}}\\ 
 0&0&{x}^{-2}\end {pmatrix} 
\mb{and}
\wt K(x)=\begin {pmatrix} 1&{\frac {{x}^{2}-1}{ \left( b+x \right) x
}}&{\frac {{x}^{2}-1}{ \left( b+x \right) x}}\\ 
0&{\frac {bx+1}{ \left( b+x \right) x}}&0\\ 
0&0&{\frac{bx+1}{ \left( b+x \right) x}}
\end {pmatrix}  .
\end{eqnarray}
One obtains
\begin{equation}
 -\frac{1}{2}K'(1)=B\mb{,}\frac{1}{2}\wh K'(1)=\wh B\mb{and}\frac{1}{2}\wt K'(1)=\wt B.
\end{equation}

We also define  the Lax operators by $\LL^{(3)}(x)=L^{(3)}(x)\wt L^{(3)}(x)$ where
\begin{eqnarray}\label{eq:lax3}
  L^{(3)}(x)=\begin{pmatrix} x+\lambda A_1 A_2&x \varepsilon_1&x \varepsilon_2\\ 
  \lambda \delta_1 A_2&\lambda  A_2&0\\ 
  \delta_2& \varepsilon_1 \delta_2&1
\end{pmatrix}\mb{and}
\wt L^{(3)}(x)= \begin{pmatrix}1& \varepsilon_3\\
0& A_3\\ 
\delta_3/x& 1/x
\end{pmatrix},
 \end{eqnarray}
 and $\LL^{(2)}(x)=L^{(2)}(x)\wt L^{(2)}(x)$ where
 \begin{eqnarray}
  L^{(2)}(x)=\begin{pmatrix} x & x\varepsilon_4\\ 
  \delta_4& 1
\end{pmatrix}\mb{and}
\wt L^{(2)}(x)= \begin{pmatrix}1\\
1/x
\end{pmatrix}.
\end{eqnarray}
Finally, we introduce the 3-component vector
\begin{equation}\label{eq:XLL}
 \boldsymbol{X}(x)=\LL^{(3)}(x)\LL^{(2)}(x).
\end{equation}

The matrices introduced in this section depend on the supplementary parameter $x$ (the 
spectral parameter).
They lead to the matrices used previously to construct the matrix ansatz by remarking that
\begin{equation}
 \boldsymbol{X}=\boldsymbol{X}(1)\mb{,}\boldsymbol{X'}=\frac{d}{dx}\boldsymbol{X}(x)\Big|_{x=1}\mb{,}L^{(3)}=L^{(3)}(1)\mb{,} L^{(3)'}=\frac{d}{dx}L^{(3)}(x)\Big|_{x=1},...
\end{equation}
As mentioned previously, the introduction of this spectral parameter makes some of our previous definitions now appear natural.
For example, the factorized form of $\LL^{(3)}(x)=L^{(3)}(x)\wt L^{(3)}(x)$ implies
\begin{equation}
 \LL^{(3)'}=\frac{d}{dx}\LL^{(3)}(x)\Big|_{x=1}=L^{(3)'}\wt L^{(3)}+L^{(3)}\wt L^{(3)'}
\end{equation}
and reproduces relation \eqref{eq:LL3}.\\

The results obtained in section \ref{sect:proof} can be deduced from the two main relations (proved in appendix \ref{GZ-ZF}):
\begin{eqnarray}
  && \check R^{(3)} (x_1/x_2) \boldsymbol{X}(x_1)\otimes \boldsymbol{X}(x_2)=\boldsymbol{X}(x_2)\otimes \boldsymbol{X}(x_1),\label{eq:ZF}\\
  && \llangle W| K(x)\,\boldsymbol{X}(1/x)= \llangle W|\boldsymbol{X}(x)
\mb{and} \overline K(x) \boldsymbol{X}(1/x)|V\rrangle = \boldsymbol{X}(x) |V\rrangle\;,
\qquad\label{eq:GZ}
\end{eqnarray}
where $\overline K(x)$ is $\wh K(x)$ or $\wt K(x)$ depending on the right boundary considered.
Taking the derivative of these relations w.r.t. $x_1$ and setting $x_1=x_2$, we recover the main relations \eqref{eq:mXX} and \eqref{eq:BXX}
used to prove the matrix ansatz.
In the context of integrable quantum field theory, an equation such as \eqref{eq:ZF} is usually called a Zamolodchikov-Faddeev (ZF) 
relation and equations \eqref{eq:GZ} are called Ghoshal-Zamolodchikov (GZ) relations. 
For more details about the use of these relations 
in the context of Markov chains see \cite{SW,CRV}.

\section{Conclusion}

In this paper, the stationary state of the open 2-species TASEP in the
case of integrable boundary conditions \eqref{P1}, \eqref{P2} is computed
using a matrix product ansatz. We find that the `matrices' are in fact
four-fold tensor products whose generators are expressed in terms of
more fundamental generators introduced to study the 1-species
TASEP. However in the stationary state there are various factorisation
properties which reduce the complexity of the calculations.  We also
show the utility of these expressions by computing exactly the
normalisation factor of the stationary state.  

The two-species models we have considered are distinguished by the
the integrability  of the boundary conditions \cite{CMRV}.
Here, we have  further shown  that
the proof of the matrix product state involves
relations that can be recovered from equations of the
Zamolodchikov-Faddeev and Ghoshal-Zamolodchikov types.

In the case of the periodic boundary conditions, a similar tensor-product construction to that 
employed here has been used for the totally asymmetric  and partially asymmetric simple 
exclusion process with N-species \cite{EFM,PEM,AAMP1,CantdeGierWheel}. The present  paper takes a  first step to generalizing these results
to the case of integrable open boundary conditions (see \cite{caley} for integrable boundaries of the N-species ASEP). 
In particular, we believe that the factorisation scheme proposed in this paper, 
using $\LL$ and $K$ matrices of decreasing sizes, see \eqref{facto1} and \eqref{eq:k1}, 
will remain valid in the N-species case  and  allow the matrix ansatz 
for an  N-species model to  be constructed from that for an (N-1)-species model.
Let us also mention that the queueing interpretation \cite{FM1,FM2,EFM,PEM,AAMP1}  provides a clear understanding  of the role of the different spaces used 
for the fundamental generators in the case of the periodic boundary conditions. 
It would be of interest to see whether a
similar analysis may also be carried out for the  open boundary case 
presented here,  which would then clarify the raison d'\^etre of each of the spaces needed in our construction.

Besides,   the matrix ansatz for the one-species ASEP was used in 
  \cite{LP}   to obtain  the Baxter's $Q$-operator of the model. We believe  
 that  the matrix ansatz found here 
may  be useful to  construct  the
$Q$-operator for the multi-species TASEP.

Finally,
  another interpretation of the matrix ansatz  was proposed recently 
  for the multi-species TASEP with periodic boundary conditions \cite{KMO}, leading to 
 new  relations between the tetrahedron equation, the stochastic $R$-matrix and the matrix ansatz. 
It would be of interest to see whether the construction  of  \cite{KMO} 
may be generalized to the cases
with integrable boundaries. In turn
  this  might reveal a 3D integrability in the matrix ansatz proposed here.

\paragraph{Acknowledgement:} 
M.E. thanks the C.N.R.S. and LAPTh for partial support during the completion of this work.

\appendix

\section{Relation between $L$ and $\wt L$ \label{app:LLt}}

In this section, we show how $\wt L$ can be obtained from $L$.

We define the transposition in the space of generators as follows
\begin{equation}
 \eps^t=\delta\mb{,}\delta^t=\eps\mb{,} A^t=A \mb{and} \llangle x|^t=|x\rrangle
\end{equation}

Let us remark that starting from an $L^{(3)}$ solution to the relation \eqref{eq:mm1}, the matrix
\begin{equation}
\overline{L}^{(3)}= U{L^{(3)}}^t U\mb{where} U=\begin{pmatrix}
                                                         0&0&1\\0&1&0\\1&0&0
                                                        \end{pmatrix}
\end{equation}
is also a solution of \eqref{eq:mm1}. We have used the following property of the local operator $m$
\begin{equation}
 U_1U_2 m_{21} U_1U_2=m\;.
\end{equation}

Starting from the realisation \eqref{eq:lax3} for $L^{(3)}$, one gets
\begin{equation}
 \overline{L}^{(3)}= \begin{pmatrix} 1&\delta_1\varepsilon_2&\varepsilon_2\\
 0&\lambda A_2& \lambda \varepsilon_1A_2\\
 \delta_2& \delta_1&1+\lambda A_1A_2
\end{pmatrix}\;.
\end{equation}
The trivial representation for the $\eps,\delta,A$ algebra is defined as $\eps=\delta=1$ and $A=0$. These values are consistent with the relation \eqref{eq:ed} and the definition of $A$.
In the $ \overline{L}^{(3)}$ matrix, we may choose the trivial representation for the generators in the space 1 (\textit{i.e.} $\varepsilon_1=\delta_1=1$ and $A_1=0$). Changing the name of space 2 to space 3 
and putting $\lambda=1$, one establishes a link with the matrix $\wt L^{(3)}$:
\begin{equation}
 \overline{L}^{(3)}\Big|_{\varepsilon_1=\delta_1=1,A_1=0,\lambda=1}= \begin{pmatrix} 1&\varepsilon_3&\varepsilon_3\\
 0&A_3&  A_3\\
 \delta_3& 1&1
\end{pmatrix} 
= \wt L^{(3)}\begin{pmatrix}
 1&0&0\\
 0&1&1
\end{pmatrix}\;.
\end{equation}
The procedure to choose the trivial representation to get a simpler matrix has been used previously for the periodic case in \cite{CantdeGierWheel}.

\section{Relation with the  algebra found in \cite{CMRV} \label{app:CMRV}}

In \cite{CMRV}, the stationary state of the 
  model $(P_2)$ with $\alpha=\frac12$ and $\beta=1$ 
 was constructed using an algebra 
 based on 9 generators $G_i$ (i=1,2,\dots,9). The algebra
 generated  by the $G_i$'s was  shown to be well defined.
 However, no explicit realisation of that algebra was given. Using
 the  matrix Ansatz found here, we give, in this appendix, 
 an explicit  representation of the $G_i$'s. 
Using the following identification \cite{CMRV}
\begin{equation} 
\boldsymbol{X}(x)=\left(\begin{array}{c}
 x^2+G_9x+G_8+G_7/x\\
 G_6x+G_5+G_4/x\\
 G_3x+G_2+G_1/x+1/x^2
 \end{array}\right)\;, 
\end{equation}
the generators $X_0$, $X_1$ and $X_2$ are given by
\begin{eqnarray}
 X_0&=& 1+ G_7+G_8+G_9\\
 X_2&=&G_4+G_5+G_6\\
 X_1&=&1+G_1+G_2+G_3\;.
\end{eqnarray}
We get the following realisation for the 9 generators $G$
\begin{eqnarray}
 &&G_1=\delta_3\varepsilon_4+\varepsilon_1\delta_2A_3+\delta_2\varepsilon_3+\delta_4\,;\quad G_2=\delta_3+\delta_2\varepsilon_4+\delta_2\varepsilon_3\delta_4+\varepsilon_1\delta_2A_3\delta_4\,;\quad G_3=\delta_2\qquad\\
 &&G_4=\lambda(\delta_1A_2\varepsilon_3+A_2A_3)\ ;\quad G_5=\lambda(\delta_1A_2\varepsilon_4+\delta_1A_2\varepsilon_3\delta_4+A_2A_3\delta_4) \ ;\quad G_6=\lambda \delta_1A_2\\
 &&G_7=\lambda A_1A_2\varepsilon_3+\varepsilon_2\quad;\quad G_8=\lambda A_1A_2\varepsilon_4+\varepsilon_2\delta_3\varepsilon_4+\lambda A_1A_2\varepsilon_3\delta_4+\varepsilon_2\delta_4+\varepsilon_1A_3+\varepsilon_3\\
 && G_9=\lambda A_1A_2+\varepsilon_2\delta_3+\varepsilon_4+\varepsilon_1A_3\delta_4+\varepsilon_3\delta_4
\end{eqnarray}
We checked using a symbolic calculation program \cite{form} that the representation presented here indeed obeys the commutation relations given in \cite{CMRV}.

 \section{Algebraic proof of relations \eqref{eq:ZF} and \eqref{eq:GZ}\label{GZ-ZF}}
We can prove relations \eqref{eq:ZF} and \eqref{eq:GZ} by direct computations. However, using the factorisation \eqref{eq:XLL},
we can split the proof of these relations into simpler ones.
\subsection{ZF relations}
One can show that the following relations hold
\begin{eqnarray}
&& \check  R^{(3)}(x_1/x_2) L^{(3)}(x_1) \otimes L^{(3)}(x_2)=L^{(3)}(x_2) \otimes L^{(3)}(x_1) \check R^{(3)}(x_1/x_2),\label{RLL1}\\
&& \check  R^{(2)}(x_1/x_2) L^{(2)}(x_1) \otimes L^{(2)}(x_2)=L^{(2)}(x_2) \otimes L^{(2)}(x_1)\check R^{(2)}(x_1/x_2),\\
&& \check R^{(3)}(x_1/x_2) \wt L^{(3)}(x_1) \otimes \wt L^{(3)}(x_2)=\wt L^{(3)}(x_2) \otimes  \wt L^{(3)}(x_1)\check R^{(2)}(x_1/x_2),\\
&& \check  R^{(2)}(x_1/x_2) \wt L^{(2)}(x_1) \otimes \wt L^{(2)}(x_2)=\wt L^{(2)}(x_2) \otimes \wt L^{(2)}(x_1),
\end{eqnarray}
where we used the braided R-matrix for the single-species TASEP, built on the local operator $m^{(2)}$  (see \eqref{markov2}): 
\beq
\check R^{(2)}(x)=1+(1-x)\, m^{(2)}.
\eeq
These identities imply
\begin{eqnarray}
&& \check  R^{(3)}(x_1/x_2) \LL^{(3)}(x_1) \otimes \LL^{(3)}(x_2)=\LL^{(3)}(x_2) \otimes \LL^{(3)}(x_1)\check  R^{(2)}(x_1/x_2), \label{RLL2}\\
&& \check  R^{(2)}(x_1/x_2) \LL^{(2)}(x_1) \otimes \LL^{(2)}(x_2)=\LL^{(2)}(x_2) \otimes \LL^{(2)}(x_1).\label{RLL3}
\end{eqnarray}
Let us remark that taking the derivative of these relations w.r.t. $x_1$ and setting $x_1=x_2$, we recover the relations we used previously.
For instance, \eqref{RLL1} implies \eqref{eq:mm1}, and \eqref{RLL2} implies \eqref{eq:tel3}. 

Finally, using \eqref{RLL2} and \eqref{RLL3}, we 
prove \eqref{eq:ZF}.

\subsection{GZ relations}
The following relations hold
\beq\label{eq:k1}
\begin{aligned}
& \llangle W|_{123} \, K(x) \LL^{(3)}(1/x) = \llangle W|_{123}\,\LL^{(3)}(x)  K^{(2)}(x) 
\mb{;} \llangle W|_{4} \, K^{(2)}(x) \LL^{(2)}(1/x) = \llangle W|_{4}\,\LL^{(2)}(x)
\qquad\\
&
\overline K(x) \LL^{(3)}(1/x)\,|V\rrangle_{123} = \LL^{(3)}(x) \overline K^{(2)}(x)\, |V\rrangle_{123} \mb{;}
\overline K^{(2)}(x) \LL^{(2)}(1/x)\,|V\rrangle_{4} = \LL^{(2)}(x) \, |V\rrangle_{4} \;,
\end{aligned}
\eeq
where we have introduced the  K-matrices for the single-species open TASEP 
\begin{eqnarray}
  K^{(2)}(x)=\begin {pmatrix}
 {x}^{2}&0\\1-{x}^{2}&1
\end{pmatrix}\mb{and}
\wb K^{(2)}(x)=\begin {pmatrix}
          1&{\frac {{x}^{2}-1}{ \left( b+x \right) x}}\\0&{\frac {bx+1}{ \left( b+x \right) x}}
         \end {pmatrix}.
\end{eqnarray}
These reflection matrices are related to the boundary matrices through
\begin{equation}
 -\frac{1}{2}\left.\frac{d}{dx}K^{(2)}(x)\right|_{x=1}=B^{(2)}\mb{,}\frac{1}{2}\left.\frac{d}{dx}\wb K^{(2)}(x)\right|_{x=1}=\wb B^{(2)}.
\end{equation}
Let us remark that taking the derivative of relations \eqref{eq:k1} w.r.t. $x$ and setting $x=1$, we recover the relations 
\eqref{eq:b1}-\eqref{eq:b2}. 

Finally, relations \eqref{eq:k1} imply equations \eqref{eq:GZ}.

\end{document}